\documentclass[%
reprint,
]{revtex4-1}

\usepackage{times,color,xcolor,amsfonts,dsfont,amssymb,amsbsy,amsthm,amsmath,algorithm,algpseudocode}
\usepackage{bm,bbm,makecell,upgreek,hyperref,multirow,gensymb,hyperref,graphics,graphicx,diagbox}

\begin{document}

	\title{Wavelength Assignment in Quantum Access Networks with Hybrid Wireless-Fiber Links}
	
	\author{Sima Bahrani, Osama Elmabrok, Guillermo Curr\'as Lorenzo, and Mohsen Razavi}
	
	\affiliation{School of Electronic and Electrical Engineering, University of Leeds, Leeds, LS2 9JT, UK}

	
	
	\begin{abstract}
		We propose a low-complexity near-optimal wavelength allocation technique for quantum key distribution access networks that rely on wavelength division multiple access. Such networks would allow users to send quantum and classical signals simultaneously on the same optical fiber infrastructure. Users can be connected to the access network via optical wireless or wired links. We account for the background noise present in the environment, as well as the Raman noise generated by classical channels, and calculate the secret key generation rate for quantum channels in the {\em finite-key} setting. This allows us to examine the feasibility of such systems in realistic scenarios when the secret key exchange needs to be achieved in a limited time scale. Our numerical results show that, by proper choice of system parameters for this noisy system, it is possible to exchange a secret key in tens of seconds. Moreover, our proposed algorithm can enhance the key rate of quantum channels, especially in high noise and/or high loss regimes of operation. 
	\end{abstract} 
	

	\maketitle
\section{Introduction}
\label{intro}

Quantum technologies are expected to lead to major advances in different fields of science and technology. This includes applications in sensing, imaging, computing, and secure communications. One of the most important applications of quantum technologies is quantum key distribution (QKD), which promises forward secrecy by relying on the laws of quantum physics. This can serve as an alternative to existing techniques for public-key cryptography, whose security relies on the computational complexity of certain mathematical problems. The widespread deployment of QKD is therefore of crucial importance, which has driven many research works in recent years. The first quantum satellite, for instance, has recently been launched to space, and the first instances of satellite-based quantum communications have been reported \cite{ChinaSatQKD,ChinaSatTelep}. Moreover, quantum networks of different sizes and topologies have already been implemented in several demonstrations \cite{secoqc, QAccess_Toshiba, Sasaki:TokyoQKD:2011, qi2010feasibility, MDInetwork, China_BeiShang}. In addition, coexistence of classical data channels with QKD channels has been demonstrated in different setups \cite{Telcordia_1550_1550, Telcordia_1550_1310,Shields.PRX.coexist,patel2014quantum,CVQKD-DWDM}. This is an enabling step to make future quantum networks cost efficient.

There are several important features that can further enrich the above developments in quantum communications networks. For instance, ease of access to quantum networks is a necessity for widespread use of QKD. This cannot necessarily be achieved by satellite, which requires large telescopes, or fiber-based, which can be inconvenient for ordinary users, connections. A {\em wireless} link is needed to connect portable QKD devices to a fiber/satellite-based quantum network. Wireless QKD links are, however, prone to loss and error. A resource efficient design is needed to optimize the users' access to a network that carries both their quantum and classical signals. Initial steps toward this objective have been taken. Experimental demonstration of QKD between a handheld device and an ATM has been achieved in \cite{HP_HandheldQKD,chun2017handheld}. Furthermore, in \cite{Elmabrok:18}, the feasibility of wireless indoor QKD has been studied. In \cite{elmabrok2018quantum}, quantum networks with wireless users connected to an access network has been considered. Optimal wavelength assignment in a hybrid quantum-classical link has also been studied \cite{bahrani2018wavelength, crosstalk2016}.

In this paper, we combine all the above features in a passive optical network (PON) where users, in addition to transmitting classical data, are enabled to exchange keys with the central office. Quantum and classical channels are multiplexed by means of dense wavelength division multiplexing (DWDM), and each user is allocated two specific wavelengths in the C-band. The users can use wireless links to connect to this DWDM-PON. For such a network, we first develop a low-complexity near-optimal technique for assigning wavelengths to quantum and classical channels. We then investigate finite-key effects in such setups \cite{Globecom2018}. This is important in at least two aspects. First, because the system works in a high-noise regime, the statistical fluctuations due to a limited block size could be severe in our system. Second, the required size of the block determines the duration of time that a wireless user must wait until the key exchange is completed. An excessive amount of delay in this process would reduce its practicality and convenience. It is then important to find out the regimes of operation that our wireless QKD network offers acceptable performance. 

While a hybrid wireless-fiber link is an attractive candidate for enabling key exchange between portable gadgets and the central office, certain issues are required to be dealt with in the process. One major problem is the background noise due to the lighting sources in the environment. Since QKD sytems rely on the transmission of weak signals, they are inherently vulnerable to such noises. In an indoor environment, however, the wireless user may be in a position to control the lighting conditions of the environment to enable QKD implementation. Another issue arises when the wireless QKD signal is collected and coupled to an optical fiber. This would introduce some coupling loss in the QKD system, which adversely affects QKD operation. To deal with such a loss, either QKD transmitter or the coupling node, or both of them, should exploit beam steering techniques. Such methods enhance the alignment between the two nodes, and substantially reduce coupling losses. In \cite{elmabrok2018quantum}, the feasibility of QKD implementation in hybrid quantum-classical access networks with wireless indoor links has been investigated. In this work, we consider a similar setup and further study the condition and regimes of operation in which wireless indoor QKD is practical.  

Apart from the aforementioned challenges, the transmission of QKD signals, which typically contain a few number of photons, alongside intense classical signals on the same optical fiber is not without its own problems. In particular, it is known that classical channels induce additional crosstalk noise on QKD channels. One major source of such a crosstalk noise is Raman scattering. The in-band noise generated by this phenomenon can be reduced, but not entirely eliminated. In particular, by using conventional techniques, e.g. spectral filtering by narrow bandpass filters (NBFs), and/or minimizing the time gate duration of detectors, one can reduce the degrading effect of this noise to some extent \cite{Shields.PRX.coexist,patel2014quantum}. More advanced techniques, such as controlling the launch power of classical channels, or the use of orthogonal frequency division multiplexing (OFDM) techniques \cite{OFDM-QKD} have also been implemented \cite{Shields.PRX.coexist}, or proposed \cite{OFDM-QKD_SPIE-Photon2016}. In \cite{bahrani2018wavelength,bahrani2016optimal}, it has been shown that another effective method is the optimal assignment of the available wavelengths to quantum and classical channels. In this paper, a new sub-optimal technique for wavelength assignment is proposed and its effect on the performance of quantum access networks is investigated. {This is particularly interesting when our access network has wireless links as this implies that the QKD system must operate in harsh conditions of high loss and background noise. In the latter scenario, optimal wavelength assignment is expected to extend the regime of operation where secure key exchange is possible. Another feature of such a hybrid access network is that the number of classical and quantum channels are identical, which further constrains the optimization problem.} 

Another important aspect of our study is the examination of finite-key effects in our hybrid system. In a typical QKD session, for example in the BB84 protocol, a certain number of qubits are transmitted. Then, some parameters, e.g. certain error probabilities, are required to be bounded to monitor the key exchange process and perform privacy amplification. The latter restricts the information leakage to a potential eavesdropper. If we send a very large set of qubits, our measured rate parameters would asymptotically be identical to the probabilities of interest. In practice, however, we have a limited time to exchange qubits, hence we have to pessimistically bound our parameters of interest based on our measurement results. This is done by introducing a failure probability parameter, $\varepsilon$, which specifies how often our pessimistic bounds are not met. Fortunately, finite-key effects in decoy-state BB84 protocol have been rigorously analyzed in several recent work. In \cite{curty2014finite}, a rigorous approach by means of Chernoff and Hoeffding inequalities is developed. This work has been extended and the bounds have been tightened in \cite{zhang2017improved}. We use the latter approach by further improving the numerical calculations in the analysis.   

The rest of the paper is organized as follows. In Sec.~\ref{sys_describe}, the system structure is described in detail. In Sec.~\ref{wave}, we consider the issue of wavelength assignment and present a low-complexity algorithm for this purpose. Section \ref{finite}, presents the finite-key analysis for the system. In Sec.~\ref{num}, our numerical results are presented, and in Sec.~\ref{con} we conclude the paper.

\section{System Description}
\label{sys_describe}
\begin{figure}[t]
	\centering
	\includegraphics[width=3 in]{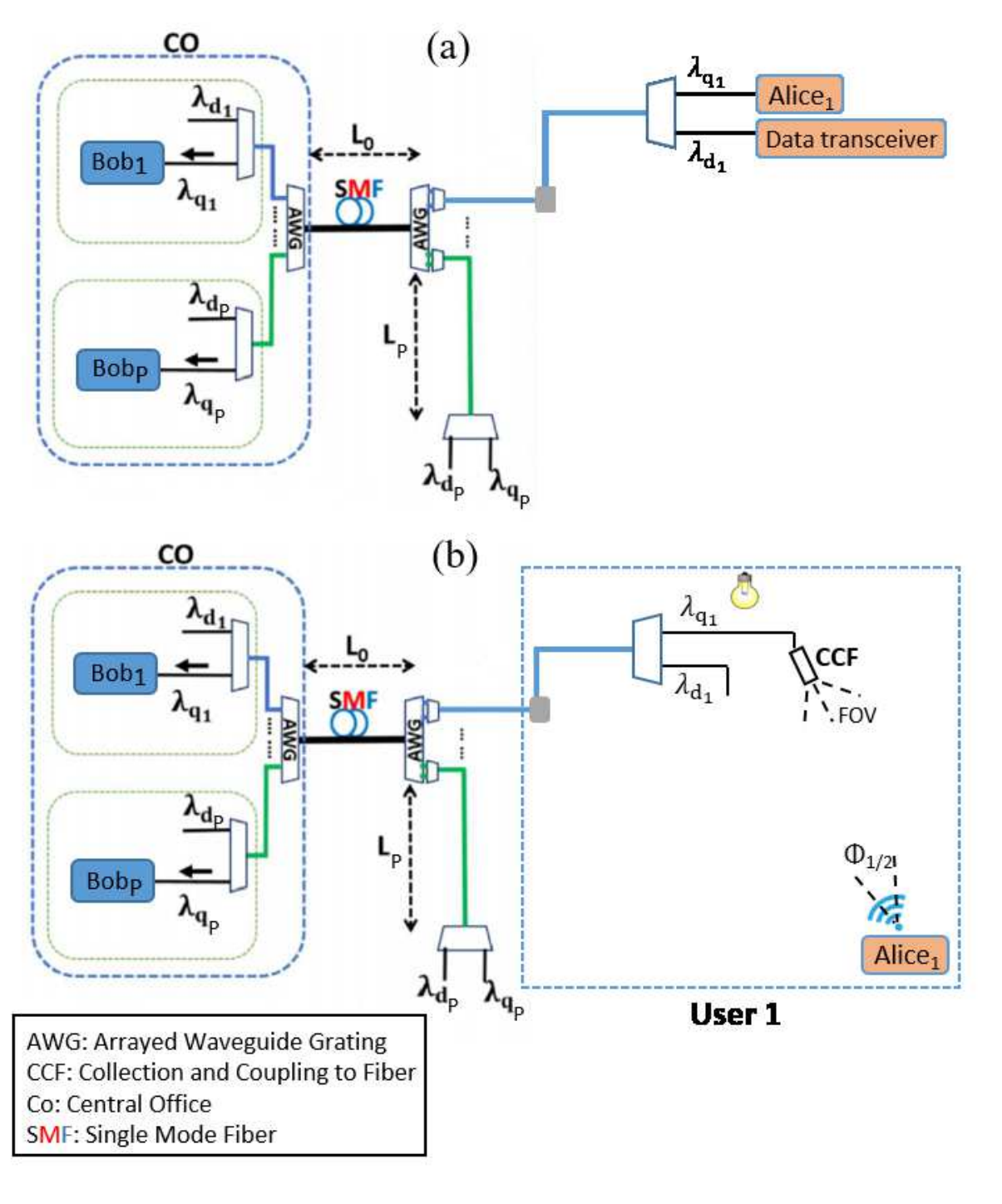}
	\caption{A quantum-classical access network in which the QKD users have either (a) directly connected to the fiber network, or (b) via embedded wireless indoor links. Each user is assigned two wavelengths, one for classical communications and one for quantum applications. In (b), the QKD signal is collected at the ceiling, using beam steering techniques, and is coupled to the fiber network. \label{system}} 
\end{figure}

In this paper, two optical access network setups for the transmission of quantum and classical signals are considered. These setups are shown in Figs.~\ref{system} (a) and (b). Both setups enable users to connect to the central office via a PON. In Fig.~\ref{system}(a), the QKD encoder is directly connected to the access fiber, whereas in Fig.~\ref{system}(b), the user is connected to the PON via wireless indoor links. In the PON structure, DWDM techniques are used to transmit classical data and weak quantum signals on the same fiber. We assume that there are $P$ users in the system, where the $i$th one is connected to the splitting point of the PON via an optical fiber of length $L_{i}$. The distance between the splitting point and the central office is denoted by $L_0$. We denote the set of quantum and classical channels by $Q=\{\lambda_{q_1},\lambda_{q_2},..., \lambda_{q_P}\}$ and $C=\{\lambda_{d_1},\lambda_{d_2},..., \lambda_{d_P}\}$, respectively. Two wavelengths $\lambda_{d_i}$ and $\lambda_{q_i}$ are assigned to the $i$th user, for $i=1,2,...,P$, corresponding to data and QKD channels, respectively. We assume that C-band is used for both quantum and classical channels. The available wavelengths in the system are represented by $G=\{\lambda_1,\lambda_2,..., \lambda_D\}$, where $D \geq 2P$. The channel spacing is denoted by $\Delta$. 

{In our setting, each user has a dedicated wavelength to transmit and receive classical data. Given that the bit rate at access networks is not as high as the backbone networks, we assume that these classical channels are bidirectional. Circulators can be used to separate uplink and downlink traffic. In our analysis, the launch power of classical signals at their input to the fiber links is denoted by $I$. Note that, in our QKD protocol, we also needs to exchange classical data between the user and the central office for post-processing purposes. The dedicated classical channel to each user can be used for this purpose too. Other required control signals, such as those needed for synchronization of QKD signals, are often exchanged at a much lower rate than the QKD pulses themselves, and are not explicitly considered in our analysis.}

In the setup shown in Fig.~\ref{system}(b), the end users are connected to the PON via wireless indoor links. In order to control the background noise generated by the light sources in the indoor environment, we assume that each user uses a windowless room with a light bulb at the center of ceiling. Although both quantum and classical setups can use wireless links, in this paper, we only consider the operation of the QKD part. We assume that a telescope is located at the ceiling to receive the wireless quantum signal and couple it to an optical fiber. Such a coupling process introduces an additional loss, denoted by $\eta_{\rm c}$, to our QKD system. One effective method to deal with this problem is to use beam steering techniques at both the QKD transmitter and the coupling node to provide full alignment between them. In this paper, we assume that the coupling loss is minimized by this method, {which can be achieved by existing adaptive tracking and pointing techniques developed for wireless optical communications. In our analysis, we assume that this initialization of the link can be done in a reasonable amount of time. We then mainly focus on the time needed to exchange QKD signals in our finite-key setting.} As for the location of the QKD transmitter, we consider the worst case scenario where it is at the corner of the room with semi-angle at half power of $\Phi_{1/2}$. 

For our QKD channels, we assume that each user is equipped with a QKD encoder and the QKD receivers are located at the central office. The QKD receiver corresponding to the $i$th user is denoted by ``${\rm Bob}_{\rm i}$" in Figs.~\ref{system} (a) and (b). It is assumed that vacuum+weak decoy-state BB84 protocol with time-bin encoding \cite{Lo:EffBB84:2005} is used at the QKD setups. We consider efficient BB84 protocol, where the probabilities of $Z$ and $X$ bases, denoted by $P_{\rm Z}$ and $P_{\rm X}$, can be asymmetric. Throughout this paper, the superscripts ``$s$'', ``$w$'', and ``$\upsilon$'', respectively, represent the signal state, weak decoy state, and vacuum state. The probabilities of choosing these states are, respectively, denoted by $q^s$, $q^w$, and $q^{\upsilon}$, where $q^s+q^w+q^{\upsilon}=1$. The intensity, i.e., the mean number of photons, of the signal and weak states are, respectively, denoted by $\mu$ and $\nu$, where $\mu > \nu$. In our work, we consider the finite-key effects, where the qubit pulses are transmitted in a limited time interval. The parameters $\mu$, $\nu$, $q^s$, $q^w$, and $P_{\rm z}$, each has a range of feasible values. We will optimize our lower bound on the final key rate over these parameters to obtain the best performance.

The copropagation of classical and weak quantum signals on the same fiber results in new challenges that should be taken care of. The most important problem is that this setting would introduce additional background noise on QKD channels. Two major sources of noise generated by classical signals are Raman scattering and adjacent channel crosstalk \cite{eraerds2010quantum}. The transmission of a classical signal in the same direction as the QKD signal generates forward Raman scattering, while such transmission in the opposite direction results in backward Raman scattering. In our setup, the uplink and downlink classical signals would introduce forward and backward Raman noise, respectively. {Both effects will be fully considered in our analysis.}

In order to reduce the crosstalk noise generated by the classical channels at the quantum receivers, different methods have been proposed in the literature \cite{Shields.PRX.coexist,patel2014quantum}. In this paper, we assume that narrow bandpass filters are used at the quantum receivers. Such filtering can suppress adjacent channel crosstalk effectively. Because of its wide bandwidth, Raman noise, however, remains a problem. In the next section, we consider optimal wavelength assignment as an effective method to reduce the deteriorating effects of this noise.
 
\section{Wavelength assignment} 
\label{wave}
In the quantum-classical access networks described in Sec.~\ref{sys_describe}, the background noise generated by Raman scattering has a deteriorating effect on the performance of QKD channels. This noise partly depends on the wavelength difference between particular quantum and classical channels. The way that we allocate the available wavelengths to our quantum and classical channels may then affect the performance of QKD systems. One possible design relies on the allocation of the lowest wavelengths to quantum channels and the longest wavelengths in the grid to classical ones. This setting is based on one of the properties of the Raman spectrum, whose magnitude is generally smaller in the anti-stokes region, as compared to the stokes region \cite{eraerds2010quantum}. We refer to this method as ``conventional method''. However, this approach may not be the optimal solution. 

In this section, we propose a low-complexity algorithm to allocate wavelengths to quantum and classical channels near-optimally, with the goal of minimizing the total sum of Raman noise at the QKD receivers. In \cite{bahrani2018wavelength}, it has been shown that for the decoy-state BB84 protocol, this goal often corresponds to the maximization of total secret key rate of quantum channels. We have verified that our near-optimal technique matches, in most realistic cases, that of an optimal solution that relies on exhaustive search as proposed in \cite{bahrani2018wavelength}. 

Let us first review the setting and the approach used in \cite{bahrani2018wavelength}. In \cite{bahrani2018wavelength}, {a single DWDM link with $N_{\rm QKD}$ quantum channels and $N_{\rm data}$ classical channels has been considered. This typically corresponds to the core of trusted-node QKD network. } A matrix-based wavelength assignment method, {which relies on exhaustive search,} has been proposed for such a DWDM link. In particular, a $D \times D$ matrix, $\bf{U}$, with elements given by
\begin{equation}
{\bf{U}}_{ij}=\left\{ 
\begin{array}{cc}
\lambda_{j}\beta(\lambda_{i},\lambda_{j})&\quad i\neq j\\
\infty&\quad i=j
\end{array}
\right.
.
\label{linear_opt_prob}
\end{equation} 
is defined. Here, $\beta(\lambda_{i},\lambda_{j})$ is the Raman cross section for a classical channel at $\lambda_i$ and a quantum channel at $\lambda_j$. Since the two classical and quantum channels have different wavelengths, we have selected $\bf{U}_{ij}=\infty$ for $i=j$. It has been shown in \cite{bahrani2018wavelength} that the problem of minimization of total sum of Raman noise on quantum channels corresponds to finding {$N_{\rm data}$} rows and {$N_{\rm QKD}$} columns of matrix $\bf{U}$ such that the sum of elements at the intersection of these rows and
columns is minimized. The algorithm proposed in \cite{bahrani2018wavelength} considers all cases and chooses the best one. {This turns out to be computationally extensive for a large number of channels or users.}

{Here, we propose a near-optimal technique for wavelength assignment, which has much lower complexity than that of \cite{bahrani2018wavelength}.} According to the numerical results presented in \cite{bahrani2018wavelength}, it can be observed that the optimal wavelength assignment has some characteristics that can be used to lower the algorithm complexity. The first one is that the resulting pattern is typically comprised of at most three separate quantum bands and three separate classical bands. This can intuitively be justified by the fact that Raman spectrum has three low-value regions \cite{bahrani2018wavelength}. Another typical feature of the optimal wavelength pattern is that the {unused} channels, in case the total number of active channels is less than the total number of channels in the grid, are next to each other, such that they make {an unused (null) band}. By considering the above features, we propose a fast and low-complexity wavelength allocation algorithm that can be used for the DWDM-PON structures in Fig.~\ref{system}. as well as the single DWDM link in \cite{bahrani2018wavelength}.

\begin{figure}[t]
 	\centering
 	\includegraphics[width=\linewidth]{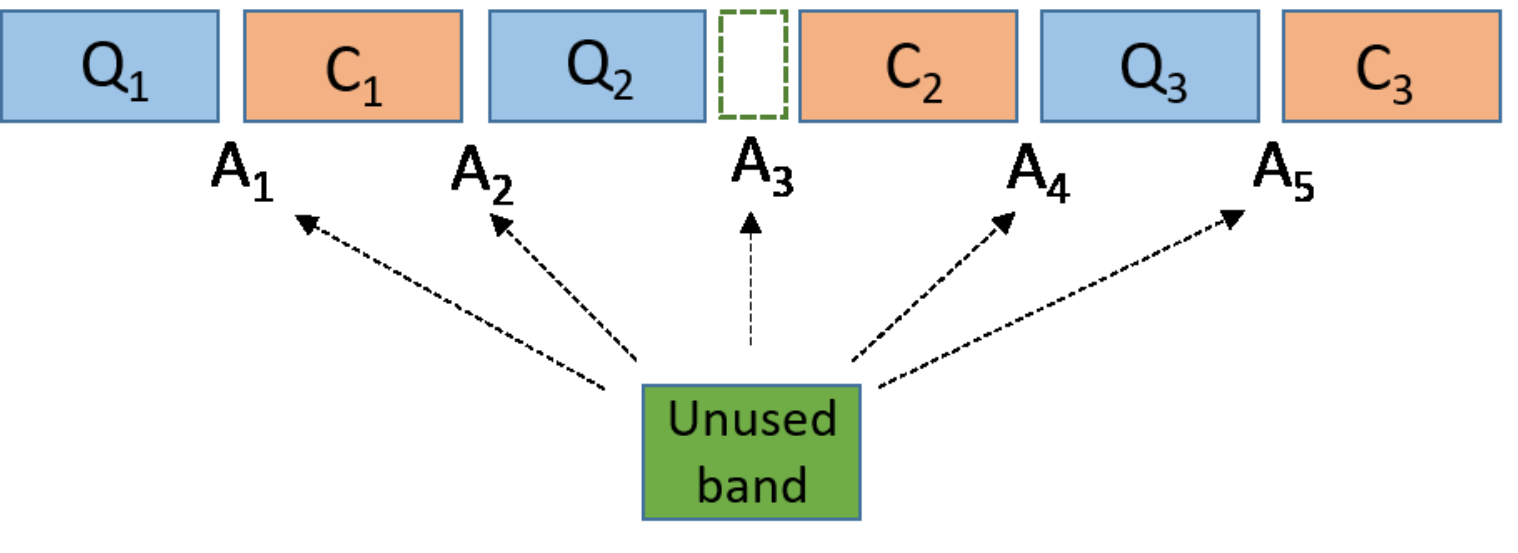}
 	\caption{Classical, quantum and {unused} bands in the proposed seven-band wavelength assignment algorithm. The {unused band} is in one of the positions labeled by $A_1$ to $A_5$ ($A_3$ in the example shown).  \label{sixband}} 
 \end{figure}

{Our seven-band near-optimal wavelength assignment algorithm, see Algorithm 1, works as follow. We assume that the final assignment is composed of three quantum bands, denoted by $\{Q_1,Q_2,Q_3\}$, and three classical bands, denoted by $\{C_1,C_2,C_3\}$, plus {an unused band} whose location can be one of the regions $A_1$, $A_2$, $A_3$, $A_4$, or $A_5$ in Fig.~\ref{sixband}. We denote the number of quantum channels in $Q_i$ by $X_i$, where $0 \leq X_i \leq N_{\rm QKD}$, and $X_1+X_2+X_3=N_{\rm QKD}$. Similarly, the number of classical channels in $C_i$ is denoted by $V_i$, where $0 \leq V_i \leq N_{\rm data}$, and $V_1+V_2+V_3=N_{\rm data}$. In our proposed algorithm, we sequentially consider all possible values for $X_i$ and $V_i$, for $i=1,\ldots,3$. We also consider five possible regions, $A_1$,..., $A_5$, for the {unused band}. By specifying a particular set of values for $X_i$ and $V_i$, for $i=1,\ldots,3$, and choosing a particular $A_j$, for $j=1,\ldots,5$, we can exactly specify the wavelengths used in the three quantum bands, $Q_1$--$Q_3$, as well as the three classical bands, $C_1$--$C_3$. In Algorithm 1, these wavelengths are specified by ${\bf q}_j$  and ${\bf c}_j$, when the unused band is $A_j$, for, respectively, quantum and classical bands. We then calculate the total Raman noise corresponding to the wavelength pattern specified by $({\bf q}_j, {\bf c}_j)$, and minimize it over $j$. By going over all possible values for $X_i$ and $V_i$, we can update this minimum setting in every round. In the end, the case that minimizes the total Raman noise on quantum channels is chosen. In Algorithm 1, this is denoted by the outout variables $\bf{q}$ and $\bf{c}$. }

In the case of the setups in Fig.~\ref{system}, we can run Algorithm 1 for $N_{\rm QKD} = N_{\rm data} = P$. But, this will only give us the set of all classical wavelengths, $\bf c$, and that of quantum ones, $\bf q$, without specifying which two wavelengths will be assigned to each user. At this stage, we can use the Hungarian method \cite{Hungarian} to assign wavelengths to quantum and classical channels of each user in an optimal way. It turns out, however, that, in our case, where {$L_0 \gg L_i$, $i=1, \ldots,P$}, this secondary optimization step would not necessarily help much. Here, we neglect the effect of optimal matching, and assume that the wavelength assigned to the $i$th user, $\lambda_{q_i}$, is specified by the $i$th element of vector ${\bf{q}}$. Similarly, $\lambda_{d_i}$ is specified by the $i$th element of ${\bf{c}}$. 

Algorithm 1 substantially reduces the complexity of finding the optimal pattern of wavelengths. For instance, in the case of access networks, the total number of cases considered in this method is given by
\begin{equation}
\kappa_1=\frac{5}{4} (P+1)^2(P+2)^2,
\end{equation}
while this parameter for the algorithm proposed in \cite{bahrani2018wavelength} is obtained by
\begin{equation}
\kappa_2= \left({D \above 0pt P}\right) .
\end{equation}
As an example, at $\Delta=0.8~{\rm nm}$ with 44 available wavelengths and 20 users, we have $\kappa_1=266805$ whereas $\kappa_2=1.761 \times 10^{12}$. This implies that the computational complexity of the proposed algorithm in this work is significantly less than the one presented in \cite{bahrani2018wavelength}. {This is particularly important if the wavelength allocation needs to be done {\em dynamically} in which case the time it takes to calculate the optimal setting would be of practical relevance. Our Algorithm 1 offers a real-time solution to this problem without necessarily sacrificing the optimality condition. In fact, our numerical results show that for a system with 200~GHz channel spacing, when Raman noise is the dominant source of noise, Algorithm 1 gives the same results as that of \cite{bahrani2018wavelength}, which relies on exhaustive search, for $N_{\rm QKD} + N_{\rm data} \leq 22$. The exhaustive search approach will become effectively intractable when the number of channels doubles, which is the case for 100-GHz channel spacing.} 

\begin{algorithm}[H]
		\caption{Seven-band wavelength assignment algorithm. {This algorithm determines the optimal location of the quantum and classical channels in the wavelength grid and return them as $\bf q$ and $\bf c$. Below, ${\bf q}_j$ and ${\bf c}_j$, for $j=1,...,5$, respectively, represent the vector of the indices of the elements of $G$, the set of all wavelengths, assigned to quantum and classical channels, assuming that the {unused band} is the region $A_j$}.}\label{opt_algorithm5}
		\begin{algorithmic}[5]
			\Statex {{\textbf{Input:}} $\bf{U}$, $N_{\rm QKD}$, $N_{\rm data}$, $D$}
			\Statex{{\textbf{Output:}}}
			\Statex {{Vector of the indices of the elements of G assigned to quantum channels, } ${\bf q}$}
			\Statex {{Vector of the indices of the elements of G assigned to classical channels, } ${\bf c}$}
			\State{$t=1000$}
			\For {$X_1=0\ldots N_{\rm QKD}$}
			\For {$X_2=0\ldots N_{\rm QKD}-X_1$}
			\For {$V_1=0\ldots N_{\rm data}$}
			\For {$V_2=0\ldots N_{\rm data}-V_1$}	
			\State {$X_3=N_{\rm QKD}-X_1-X_2$}
			\State {$V_3=N_{\rm data}-V_1-V_2$}
			\State {${\bf q}_1=[1:X_1,D-(V_3+X_3+V_2+X_2)+1:D-(V_3+X_3+V_2),D-(V_3+X_3)+1:D-V_3]$}
			\State {${\bf c}_1=[D-(V_3+X_3+V_2+X_2+V_1)+1:D-(V_3+X_3+V_2+X_2),D-(V_3+X_3+V_2)+1:D-(V_3+X_3),D-V_3+1:D]$}
			\State {${\bf q}_2=[1:X_1,D-(V_3+X_3+V_2+X_2)+1:D-(V_3+X_3+V_2),D-(V_3+X_3)+1:D-V_3]$}
			\State {${\bf c}_2=[X_1+1:X_1+V_1,D-(V_3+X_3+V_2)+1:D-(V_3+X_3),D-V_3+1:D]$}
			\State {${\bf q}_3=[1:X_1,X_1+V_1+1:X_1+V_1+X_2,D-(V_3+X_3)+1:D-V_3]$}
			\State {${\bf c}_3=[X_1+1:X_1+V_1,D-(V_3+X_3+V_2)+1:D-(V_3+X_3),D-V_3+1:D]$}
			\State {${\bf q}_4=[1:X_1,X_1+V_1+1:X_1+V_1+X_2,D-(V_3+X_3)+1:D-V_3]$}
			\State {${\bf c}_4=[X_1+1:X_1+V_1,(X_1+V_1+X_2)+1:(X_1+V_1+X_2)+V_2,D-V_3+1:D]$}
			\State {${\bf q}_5=[1:X_1,X_1+V_1+1:X_1+V_1+X_2,X_1+V_1+X_2+V_2+1:X_1+V_1+X_2+V_2+X_3]$}
			\State {${\bf c}_5=[X_1+1:X_1+V_1,(X_1+V_1+X_2)+1:(X_1+V_1+X_2)+V_2,D-V_3+1:D]$}
			\For {$i=1: 5$}
			\State {${\bf{Z_i}}={\bf{U}}({\bf q_i},{\bf c_i})$}
			\State {${\bf s}(i)=\sum_{k=1}^{N_{\rm QKD}}{\sum_{l=1}^{N_{\rm data}}{{\bf{Z_i}}(k,l)}}$}
			\EndFor
			\State {$[{d},{\rm index}]=\min({\bf s})$}
						\If {$d<t \;$}
			\State {$t=d$}
			\State {${\bf q}={\bf q}_{\rm index}$}
			\State {${\bf c}={\bf c}_{\rm index}$}
			\EndIf
			\EndFor
			\EndFor
			\EndFor
			\EndFor
		\end{algorithmic}
\end{algorithm}  

\section{Finite-Key Analysis}
\label{finite}

The security of BB84 protocol relies on quantifying the information leakage to a potential eavesdropper. This task is performed by bounding relevant single-photon parameters, e.g., yield of single photons and their error probability. In practice, these probabilities are estimated by calculating the corresponding rates obtained from the measurement results in a QKD experiment. With the assumption of an infinitely large data size, the estimation error would converge to zero. However, in a real scenario where the data size is finite, the target probabilities and their corresponding rates may not be equal. Hence, in order to reliably generate secret keys in a finite-key setting, such statistical fluctuations should rigorously be considered.

In this section, the finite-key effects for the QKD setups in the system described in Sec.~\ref{sys_describe} are investigated. {In \cite{Globecom2018}, the basic framework for the finite-key analysis of the wireless QKD setup in Fig.~\ref{system}(b) has been developed. Here, we summarize the framework in \cite{Globecom2018}, and provide more detail to the analysis so that we can employ it for other use cases that we consider in this paper. That includes both Figs.~\ref{system}(a) and (b), with and without the near-optimal wavelength allocation technique proposed here.} 

Based on the GLLP analysis for the BB84 protocol presented in \cite{GLLP_04}, the final key size extracted from key bits in basis $\zeta$ is lower bounded by
\begin{equation}
K^{\zeta} \geq M_{1}^{s\zeta} [1-h(e_{1}^{ps\zeta})]-f M^{s\zeta} h(E^{s\zeta}),
\label{key}
\end{equation}
where $\zeta$ is either $Z$ or $X$, and $f \geq 1$ represents the error correction inefficiency. Here, $h(p)=-p {\rm log}_{2}(p)-(1-p) {\rm log}_{2}(1-p)$ is the Shannon binary entropy function. Furthermore, $M^{s\zeta}$, $E^{s\zeta}$, $M_{1}^{s\zeta}$, and $e_{1}^{ps\zeta}$, respectively, denote the number of successful detection events, the quantum bit error rate, the number of successful detection events from single-photon components, and the phase error rate of single-photon components all for the signal ($s$) state in basis $\zeta$. The first two parameters can directly be measured in experiment. However, the single-photon parameters should be bounded carefully. These bounds are then used in the privacy amplification step. In the decoy-state BB84 protocol, a lower bound on $M_{1}^{s\zeta}$ and an upper bound on $e_{1}^{ps\zeta}$ can be obtained by the use of decoy states. 

In the vacuum+weak decoy-state protocol, two decoy states are used: weak decoy state, and vacuum decoy state. This would enable us to obtain a set of observed parameters corresponding to signal and decoy states. In \cite{zhang2017improved}, the set of observables $A=\{M^{s\zeta},E^{s\zeta}M^{s\zeta},M^{w\zeta},E^{w\zeta}M^{w\zeta},M^{\upsilon\zeta},E^{\upsilon\zeta}M^{\upsilon\zeta}\}$ are used to calculate the required bounds rigorously. In our work, we have used this method to analyze the finite-key effects in the system. In the following, the main steps of this technique are outlined. 

First of all, we bound the average of each observable in $A$ by using the Chernoff bound. For any $\chi \in A$, its average is represented by ${\rm E}[\chi]$. We can then find a lower bound on ${\rm E}[\chi]$, denoted by ${\rm E}^{L}[\chi]$, and an upper bound on ${\rm E}[\chi]$, denoted by ${\rm E}^{U}[\chi]$, such that $\Pr\{{\rm E}^{L}[\chi]< {\rm E}[\chi] < {\rm E}^{U}[\chi]\} \geq 1- \varepsilon$. Here, $\varepsilon$ represents the failure probability for this bounding step. In \cite{zhang2017improved}, for an observable $\chi >0$, the following bounds have been derived:
\begin{equation}
\label{Eq:EL}
{\rm E}^{L}[\chi]=\frac{\chi}{1+{\delta}^{L}}
\end{equation} 
\begin{equation}
{\rm E}^{U}[\chi]=\frac{\chi}{1-{\delta}^{U}},
\end{equation} 
where ${\delta}^{L}$ and ${\delta}^{U}$ can be obtained by solving the following two equations:
\begin{equation}
\left(\frac{e^{{\delta}^{L}}}{(1+{\delta}^{L})^{(1+{\delta}^{L})}}\right)^\frac{\chi}{1+{\delta}^{L}}=\frac{\varepsilon}{2}
\end{equation} 
\begin{equation}
\left(\frac{e^{{-\delta}^{U}}}{(1-{\delta}^{U})^{(1-{\delta}^{U})}}\right)^\frac{\chi}{1-{\delta}^{U}}=\frac{\varepsilon}{2}.
\end{equation}
{In \cite{zhang2017improved}, the above equations are solved numerically. Here, we solve these equations analytically} using the Lambert $W$ function. We find that
\begin{equation}
\frac{1}{1+{\delta}^{L}}=-W_0(-e^{\frac{{\rm ln}(\varepsilon/2)-\chi}{\chi}})
\end{equation}
and
\begin{equation}
\frac{1}{1-{\delta}^{U}}=-W_{-1}(-e^{\frac{{\rm ln}(\varepsilon/2)-\chi}{\chi}}).
\end{equation}
If $\chi =0$, the bounds are simply given by ${\rm E}^{L}[\chi]=0$ and ${\rm E}^{U}[\chi]=-{\rm ln}(\varepsilon/2)$. These bounds can, then, be used to calculate a lower bound on $M_{1}^{\zeta}$, denoted by $M_{1}^{\zeta L}$ and an upper bound on $e_{1}^{b\zeta}$, denoted by $e_{1}^{b\zeta U}$, where $e_{1}^{b\zeta}$ is the bit error rate of single-photon components in basis $\zeta$. The parameters $M_{1}^{\zeta L}$ and $e_{1}^{b\zeta U}$ are given by \cite{zhang2017improved}
\begin{equation}
M_{1}^{\zeta L}=Y_{1}^{\zeta L} N^{\zeta}(e^{-\mu}\mu q^s+e^{-\nu}\nu q^w),
\end{equation}
\begin{equation}
e_{1}^{b\zeta U}=\frac{\frac{{\rm E}^{U}[E^{w\zeta}M^{w\zeta}]}{q^w N^{\zeta}}e^{\nu}-\frac{{\rm E}^{L}[E^{\upsilon\zeta}M^{\upsilon\zeta}]}{q^{\upsilon}N^{\zeta}}}{Y_{1}^{\zeta L}\nu},
\end{equation}
where 
\begin{eqnarray}
Y_{1}^{\zeta L}=\frac{\mu}{\mu\nu-\nu^2}\left((\frac{{\rm E}^{L}[M^{w\zeta}]}{q^w N^{\zeta}})e^{\nu}-(\frac{{\rm E}^{U}[M^{s\zeta}]}{q^s N^{\zeta}})e^{\mu}\frac{\nu^2}{\mu^2}\right.\nonumber\\
\left.-(\frac{{\rm E}^{U}[M^{\upsilon\zeta}]}{q^{\upsilon} N^{\zeta}})\frac{\mu^2-\nu^2}{\mu^2} \right).
\label{Eq:Y1}
\end{eqnarray}  
Here, $N^{\zeta}=P_{\zeta}^2 N$, where $P_{\zeta}$ represents the probability of choosing basis $\zeta$ at either transmitter or receiver and $N$ is {the total number of transmitted pulses} in a QKD round.

In the next step, $M_{1}^{\zeta L}$ is used to obtain a lower bound on $M_{1}^{s\zeta}$, denoted by $M_{1}^{s \zeta L}$. Defining $p_{1}^{s\zeta}$ as the conditional probability that a single-photon component belongs to signal state, we can write ${\rm E}[M_{1}^{s\zeta}]=p_{1}^{s\zeta}M_{1}^{\zeta}$. Then, by using the symmetric form of the Chernoff bound for the parameter $\bar{\chi}=p_{1}^{s\zeta}M_{1}^{\zeta L}$, $M_{1}^{s \zeta L}$ can be calculated. 

Finally, an upper bound on $e_{1}^{ps \zeta}$ is derived. We can apply the random sampling method to obtain $e_{1}^{psz U}$ from $e_{1}^{bxU}$. Similarly, $e_{1}^{psx U}$ is calculated from $e_{1}^{bzU}$. In the end, the parameters $M_{1}^{s \zeta L}$, $M_{1}^{s \zeta L}$, $e_{1}^{psz U}$, and $e_{1}^{psz U}$ are used in (\ref{key}) to obtain $K^x$ and $K^{z}$. The final key size extracted from both bases would, then, be given by $K=K^x+K^z$. For more details on the finite key analysis of decoy-state BB84 protocol, please refer to \cite{zhang2017improved}.

In the finite-key analysis presented in this section, we have a set of free parameters: $\mu$, $\nu$, $q^s$, $q^w$, and $P_Z$. To obtain the best performance, we optimize the key rate over possible range of these values. This requires us to solve a multivariate optimization problem. To this aim, we consider a set of initial values and find the best choice for each parameter, assuming the other parameters are constant. This process should be continued until the key rate converges to a specific value with a desired accuracy. 

\section{Numerical Results}
\label{num}

\begin{table}[t]
	\caption{Nominal values used for system parameters.}
	\centering 
	\begin{tabular}{|c |c |} 
		\hline
		Parameter & Value\\
		\hline
		Quantum Efficiency & 0.3\\
		Receiver dark count rate & 1E-6 ${\rm ns}^{-1}$\\
		Error correction inefficiency, $f$& 1.22\\
		Misalignment probability, $e_d$& {0.033}\\
		Detector gate interval and pulse width& 100 ps\\
		Fiber attenuation coefficient & 0.2 dB/km\\
		AWG insertion loss & 2 dB\\
		Repetition rate of QKD setup & 1 GHz\\
		Bandwidth of NBF & 25 GHz\\
		Failure probability, $\varepsilon$& $10^{-10}$\\ 
		\hline
	\end{tabular}
	\label{Tab:Par}
\end{table}

In this section, we evaluate the effects of wavelength assignment and finite-key setting in the systems described in Sec.~\ref{sys_describe}. To this end, we consider various numerical examples and examine the performance of the system in the finite-key regime for both the conventional method and the near-optimal wavelength assignment of Algorithm 1. A DWDM-PON structure with $L_0=5~{\rm km}$ and $L_k=500~{\rm m}$, for $k=1,...,P$, is considered. We assume that the wavelengths in the range of $1530~{\rm nm}$ to $1564.4~{\rm nm}$ are used in the system. As for the channel spacing, we consider the two cases of $\Delta=0.8~{\rm nm}$ (100 GHz) and $\Delta=1.6~{\rm nm}$ (200 GHz), corresponding to $D=44$ and $D=22$ channels, respectively. In the conventional wavelength assignment method, the wavelengths $\lambda_{q_1}=1530~{\rm nm}$ to $\lambda_{q_P}=1530~{\rm nm}+\Delta(P-1)$ are used for QKD, and $\lambda_{d_1}=1564.4~{\rm nm}-\Delta(P-1)$ to $\lambda_{d_P}=1564.4~{\rm nm}$ are used for classical signals. In the case of using the near-optimal wavelength assignment method, the wavelengths assigned to quantum and classical channels are determined by Algorithm 1.

For the indoor wireless system shown in Fig.~\ref{system}(b), it is assumed that the QKD transmitter is located at the corner of the room, and full alignment exists between the QKD transmitter and the telescope. The field of view (FOV) of the telescope and the semi-angle at half power of the QKD source are assumed to be $6^{\circ}$ and $1^{\circ}$, respectively. Nominal values for system parameters relevant to the indoor environment are chosen similar to the ones presented in \cite{elmabrok2018quantum}. Other system parameters and their nominal values are listed in Table~\ref{Tab:Par}. These parameters have been chosen based on practical considerations. 

In order to analyze the finite-key effects in our system, the observable parameters in the set $A$ are required to be quantified. In reality, these parameters are obtained during a single QKD run when $N$ signals are transmitted from Alice to Bob. Here, we assume that the measured values for these parameters are equal to the values that can be obtained analytically in the asymptotic limit scenario, when no eavesdropper is present. To calculate these values, we use the key rate analysis presented in equations (35)-(36) in \cite{zhang2017improved}. Furthermore, our calculations for the Raman noise and bulb noise (for indoor wireless system) are based on the analysis in \cite{elmabrok2018quantum}.   

\begin{figure}[t]
	\centering
	\includegraphics[width=\linewidth]{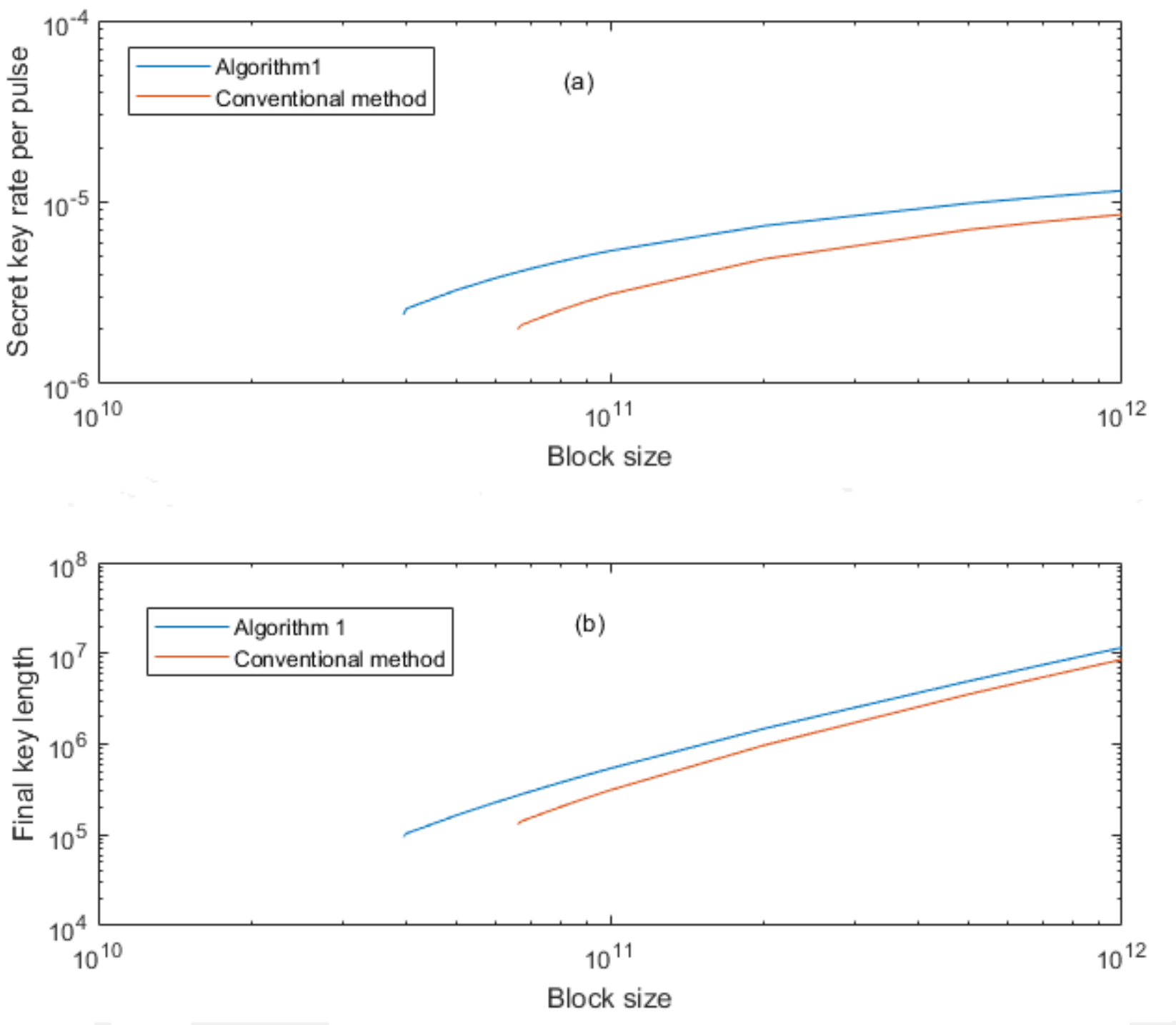}
	\caption{(a) Secret key rate per pulse at different values of block size for the channel with maximum background noise for the setup of Fig.~\ref{system}(b). (b) Final key length for different values of block size. Here, $P=20$, $\eta_{\rm c}=16~{\rm dB}$, $\Delta = 0.8~{\rm nm}$, and $I=-30~{\rm dBm}$. Other parameter values are listed in Table~\ref{Tab:Par}. \label{fig_block}} 
\end{figure} 

\begin{figure}[t]
	\centering
	\includegraphics[width=\linewidth]{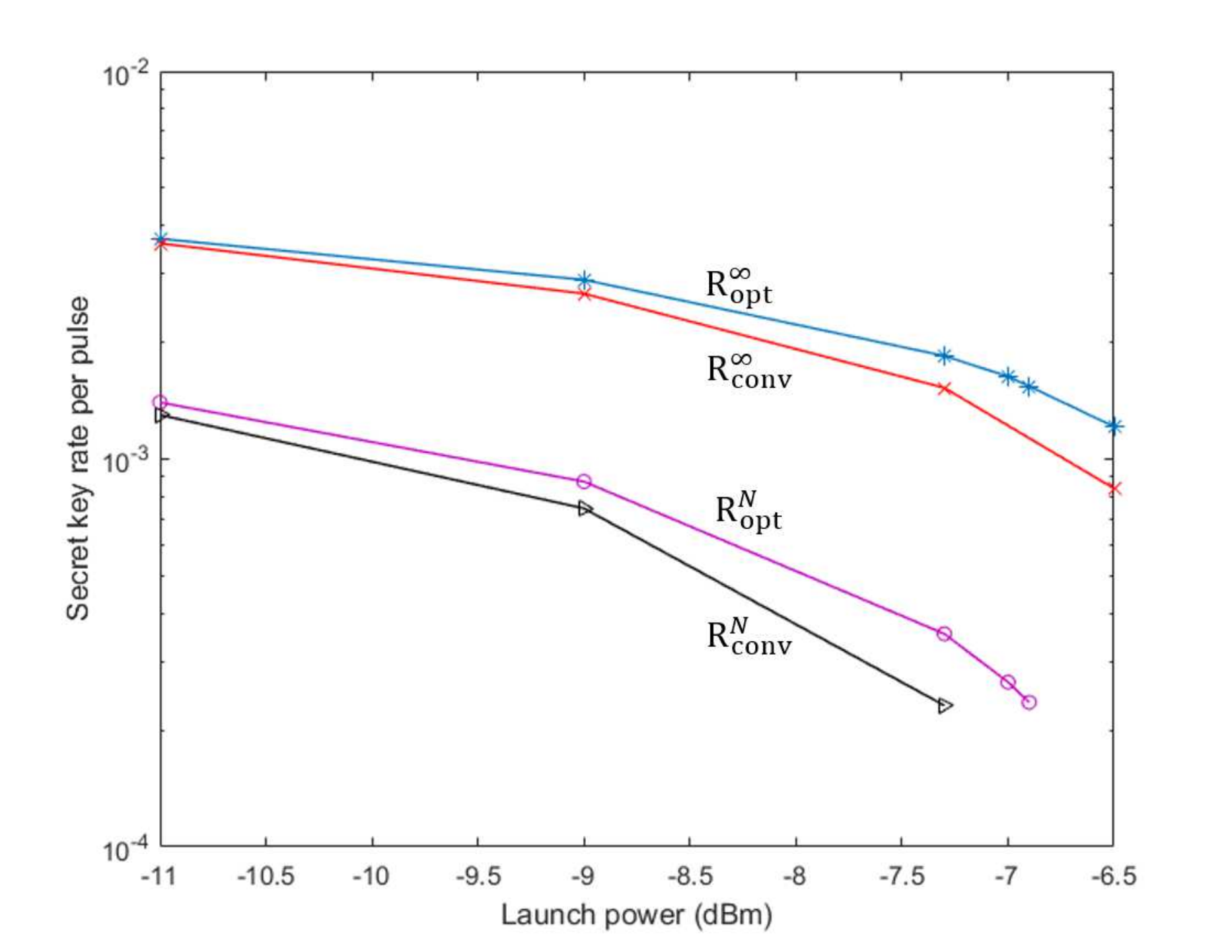}
	\caption{Average secret key rate in finite-key and asymptotic cases at different values of launch power for the setup of Fig.~\ref{system}(a). The number of users is 6, $\Delta=1.6~{\rm nm}$, and $N= 10^{10}$. {The simulations have been performed at the points represented by ``$\ast$", ``$\times$", ``$\circ$", or ``$\rhd$".   } \label{fig_power_fiber}} 
\end{figure}

The main figure of merit used in our analysis is the secret key generation rate for QKD users. In our optimization problem, we require that all $P$ QKD users have positive key rates, otherwise our effective number of QKD users would be less than $P$. That said, given that different users can be exposed to different levels of background noise, their secret key generation rates are not necessarily identical. Two particular rates would be of interest then: the one corresponding to the worst-case scenario, i.e., that of the channel with minimum rate, and the average rate of all users. It turns out that the two measures are not far from each other and that reaffirms similar results reported in \cite{bahrani2018wavelength}. We will then use either of these rates in the forthcoming graphs. As a matter of notation, the average secret key rate of users in the finite-key regime, for Algorithm 1 and the conventional method, are, respectively, denoted by $R_{\rm opt}^{N}$ and $R_{\rm conv}^{N}$. Moreover, the relative difference between these two rates, as a criteria for the rate enhancement that we obtain by applying Algorithm 1 instead of the conventional method, is represented by $\Gamma^{N}$. This parameter is given by
\begin{equation}
	\Gamma^{N}=\frac{R_{\rm opt}^{N}-R_{\rm conv}^{N}}{R_{\rm conv}^{N}}
\end{equation} 
In a similar way, the average secret key rate of users in the asymptotic limit of an infinitely long key, for Algorithm 1 and the conventional method, are, respectively, denoted by $R_{\rm opt}^{\infty}$ and $R_{\rm conv}^{\infty}$, with the relative difference between them being denoted by $\Gamma^{\infty}$. 

In the following, we study the effect of different system parameters and how our proposed wavelength assignment can improve the performance.
 
\subsection{Block size and running time}

We first look at the running time of the protocol in the wireless setup \cite{Globecom2018}. For a wireless QKD setup, this time parameter is of practical importance, because it would be inconvenient for the users if they have to wait an unreasonably long time to exchange some secret key bits. The running time primarily depends on the block size, as well as the pulse rate of the system. This implies that for a limited time interval and a fixed pulse rate, we have a limitation on our block size. Here, {we do not account for the time that it takes for the system to set up (including for initial beam alignment)} as well as that of post-processing, which can possibly be done off-line.

To find out the time requirements on the end users, we look at an extreme point, where our quantum access network is working near its full capacity with 20 users. Note that each user in our setup requires two wavelengths, so that the total number of employed wavelengths is 40 out of 44 for 100~GHz channel spacing. The parameters $\eta_{\rm c}$ and $I$ are assumed to be $16~{\rm dB}$ and $-30~{\rm dBm}$, respectively. In this scenario, we consider the worst channel, which tolerates the highest crosstalk noise. The secret key rate of this channel, at different values of block size, for both conventional and near-optimal wavelength assignment methods, are shown in Fig.~\ref{fig_block}(a). The final key size, i.e., the product of the block size and the key rate, is also depicted in Fig.~\ref{fig_block}(b). The first point to notice in Fig.~\ref{fig_block}(a) is the existence of a minimum block size below which at least one of the users would not be able to exchange a secret key. This block size for the conventional assignment is around $N=7 \times 10^{10}$, whereas for our Algorithm 1 is given by $N=4 \times 10^{10}$. For a pulse rate of 1~GHz, these values, respectively, correspond to 70~s and 40~s. There is obviously an advantage in using the near-optimal algorithm. That said, the minimum required block size in this scenario is orders of magnitude higher than what we typically need in a fiber-based link. The key reason for that is the amount of noise present in the wireless system, which makes the achievable quantum bit error rate (QBER) far from zero. For large values of QBER, we have little room for loose bounds on the error terms, which requires rather large block sizes to distill a secret key. A minute or two for wireless key exchange could still be acceptable if one compares it with the amount of time one may spend at an ATM machine. One can calculate, from Fig.~\ref{fig_block}(b), what block size, or its corresponding running time, is needed to obtain a certain number of secret key bits. The larger the size of the final key more efficient the protocol becomes as we get closer to the asymptotic regime. For the rest of this paper, unless otherwise noted, we assume a block size of $N=10^{11}$, which corresponds to 100~s at 1~GHz pulse rate. At $N=10^{11}$, in Fig.~\ref{fig_block}(b), our proposed algorithm offers nearly two times longer a key than the conventional technique.  




\subsection{Launch power and number of users}

In order to study the effectiveness of our wavelength assignment technique, here, we consider two parameters that directly affect the amount of {background} noise: launch power of data channels, $I$, and the number of users, $P$.  An increase in $P$ corresponds to an increase in the total Raman noise generated by classical users. Also, since the power of Raman noise is proportional to $I$, any increase in $I$ would again result in larger amount of Raman noise. We should then in principle use the lowest acceptable launch power that guarantees a target bit error rate for data channels. In our setup, we assume that this minimum acceptable launch power for an error rate of $10^{-9}$ is below $-30~{\rm dBm}$. This is, however, not a typical regime of operation for classical optical communications as often the launch power could be as high as 0~dBm. It would be interesting to find out what levels of power, in our access networks of Fig.~\ref{system}, would allow for the coexistence of classical and quantum channels, as we discuss next.
     
\begin{figure}[t]
	\centering
	\includegraphics[width=\linewidth]{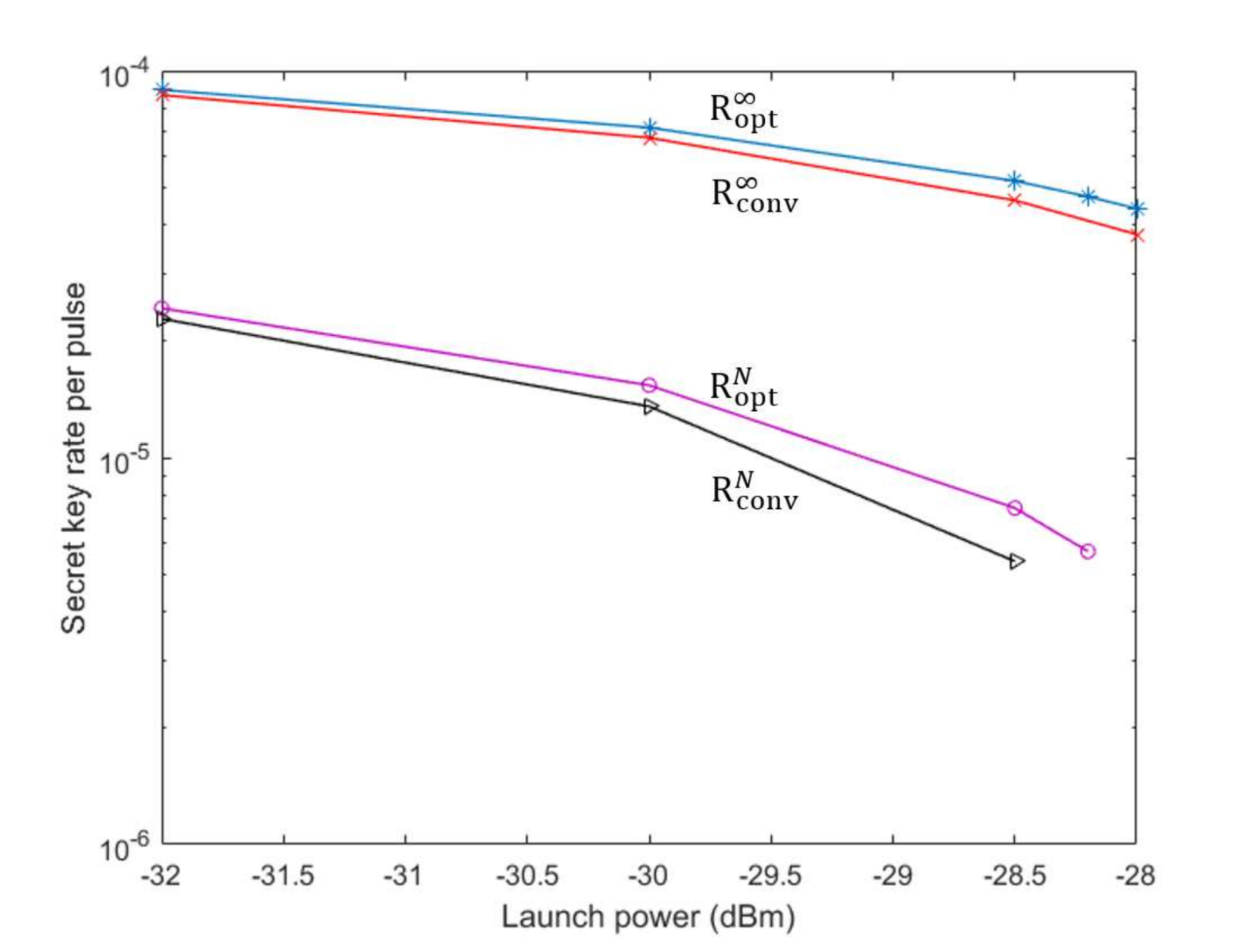}
	\caption{Average secret key rate in finite-key and asymptotic cases at different values of launch power for the setup of Fig.~\ref{system}(b). Here, $P=15$, {$\eta_{\rm c}=16~{\rm dB}$}, $\Delta=0.8~{\rm nm}$, and $N= 10^{11}$. {The simulations have been performed at the points represented by ``$\ast$", ``$\times$", ``$\circ$", or ``$\rhd$". The free parameters have been optimized for each individual channel at each point. Table~\ref{Tab:Par3} provides detailed information about these optimal values for the channel with highest crosstalk noise in the near-optimal setting.}   \label{fig_power}} 
\end{figure} 

\begin{table}[b]
	\caption{Optimal values of decoy-state BB84 parameters for different values of launch power, at the channel with highest crosstalk noise, in the near-optimal wavelength assignment setting. Here, $P=15$, $\eta_{\rm c}=16~{\rm dB}$, $\Delta=0.8~{\rm nm}$, and $N= 10^{11}$. We observe a trend in which for lower levels of background noise $q^s$ and $P_Z$ increase, while $q^w$ decreases. At high levels of noise, corresponding to larger values of $I$, $P_Z$ appraoches 0.5.}
	\centering 
	\begin{tabular}{|c |c |c|c|c|c|} 
		\hline
		$I$ (dBm)& $\mu$&$\nu$&$q^s$&$q^w$& $P_Z$\\
		\hline
-32 & 0.407 &  0.09& 0.835& 0.112 & 0.68 \\
-30 & 0.394 & 0.096 & 0.778&0.139&0.5 \\
-28.5 & 0.386 & 0.091 & 0.628&0.227&0.5 \\
-28.2 & 0.384 & 0.089 & 0.566&0.255&0.5 \\
		\hline
	\end{tabular}
	\label{Tab:Par3}
\end{table}

Let us first focus on Fig.~\ref{system}(a) with a rather low number of users $P=6$. This is a fully fiber-based access network, which can be implemented with today's technologies. Compared to the wireless indoor system of Fig.~\ref{system}(b), this setup is required to tolerate much less background noise and loss, since the noise of the lighting sources in the environment and the coupling loss are not present in this system. Moreover, because of the low number of users, we can use a larger channel spacing of $\Delta = 1.6$~nm. Given that the number of users is rather low, we may expect that this system can tolerate high values of launch power. As shown in Fig.~\ref{fig_power_fiber}, it turns out, however, that, at $N= 10^{10}$, the maximum launch power tolerated by QKD channels is around -7~dBm, which is lower than the typical value of 0~dBm. This shows the importance of power control even in seemingly simple scenarios. Our near optimal technique can roughly buy us an extra 0.5~dBm in terms of power margin. Nevertheless, at $I=-7.3~{\rm dBm}$, we have a rate enhancement of $\Gamma^{N}=53 \%$ as compared to the conventional assignment. 

Note that here we look at the average key rate and we require that all 6 users have positive key rates. The end point on each curve would then correspond to the case where one user is unable to exchange secret keys. The typical cliff-edge decline of the key rate to zero would happen later when the key rate becomes zero for all users (not shown). This somehow also justifies why the extent of improvement from our near-optimal assignment technique is on the order of tens of percents. As shown in \cite{bahrani2018wavelength}, once we impose the condition that all users have positive key rates, we enforce the system to work in its linear regime where error rates are well below the cut-off threshold for QKD systems. Optimal wavelength assignment would then offer a moderate advantage over the conventional method. 
 
\begin{figure}[t]
	\centering
	\includegraphics[width=\linewidth]{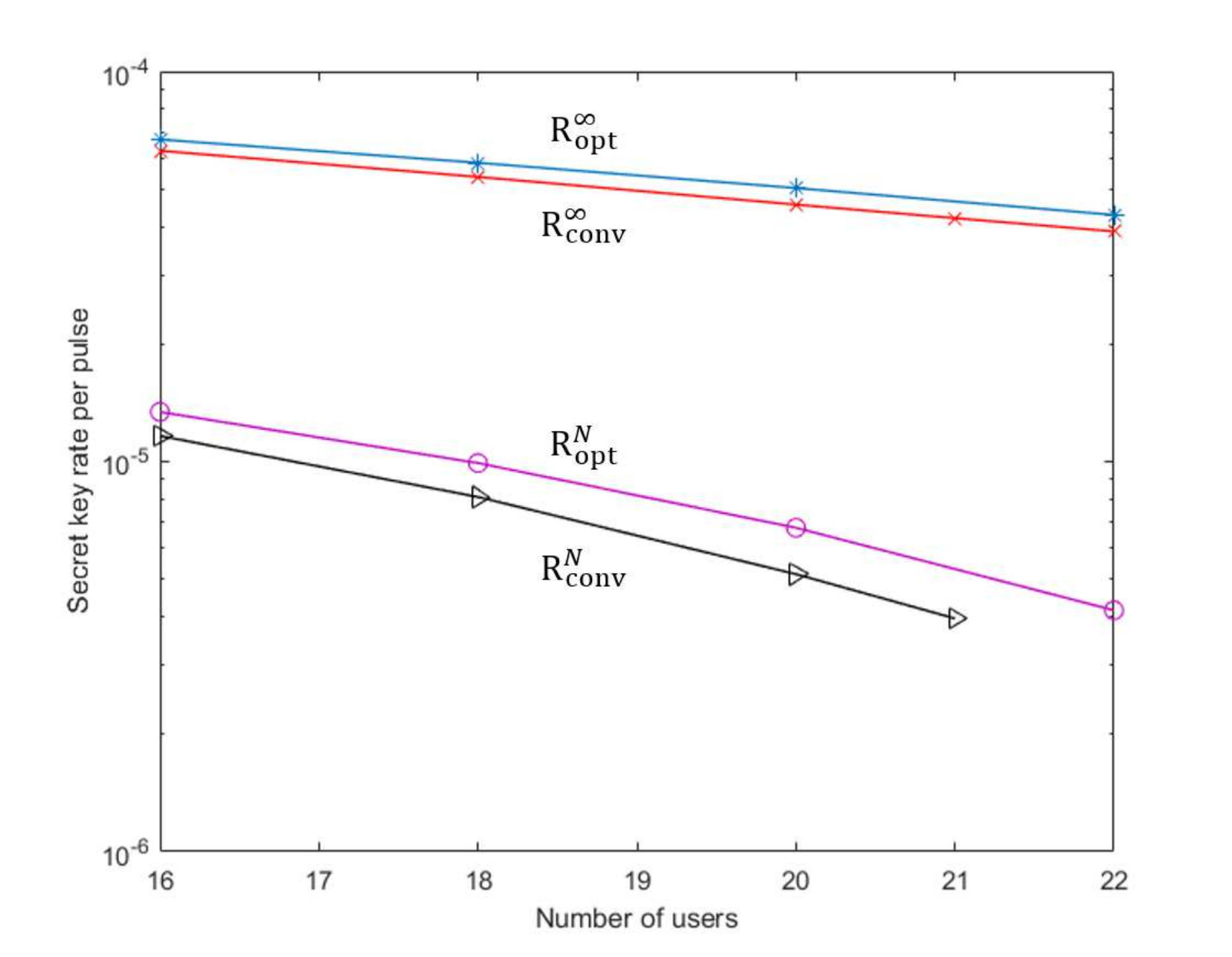}
	\caption{Secret key rate in finite-key and asymptotic cases at different number of users for the setup of Fig.~\ref{system}(b). Here, $\eta_{\rm c}=16~{\rm dB}$, $I=-30~{\rm dBm}$, $\Delta=0.8~{\rm nm}$, and $N= 10^{11}$. {The simulations have been performed at the points represented by ``$\ast$", ``$\times$", ``$\circ$", or ``$\rhd$". The free parameters have been optimized for each individual channel at each point. As an example, at $P=22$, for the channel with the highest crosstalk noise in the optimal setting, we have $\mu=0.386$, $\nu=0.091$, $q^s=0.606$, $q^w=0.236$, and $P_Z=0.5$.   }   \label{fig_users}} 
\end{figure}

\begin{table}[b]
	\caption{Maximum possible launch power (in dBm) for different values of $N$ and $P$ for the setup of Fig.~\ref{system}(b). Here, $\eta_{\rm c}=16~{\rm dB}$, and $\Delta=0.8~{\rm nm}$.}
	\centering 
	\begin{tabular}{|c |c |c|c|c|} 
		\hline
		\backslashbox{N}{P} & 15&17&19&21\\
		\hline
		$10^{10}$& -30.7& -31.3& -31.9&-32.2\\
		$10^{11}$& -28.2& -28.8& -29.3&-29.6\\
		$10^{12}$& -27.6& -28.2& -28.7&-29\\
		\hline
	\end{tabular}
	\label{Tab:Par2}
\end{table}

Figure \ref{fig_power} shows the other extreme when we are using the setup of Fig.~\ref{system}(b) with a rather high number of users $P=15$ at $\Delta = 0.8$~nm. As can be seen the maximum amount of launch power is now much lower at around {-28~dBm}. The gain in the power margin is again low, around 0.3~dBm, for the near-optimal assignment, but, at {$-28.5~{\rm dBm}$}, we achieve a rate enhancement of about $\Gamma^{N}=37 \%$.


\begin{figure}[t]
	\centering
	\includegraphics[width=\linewidth]{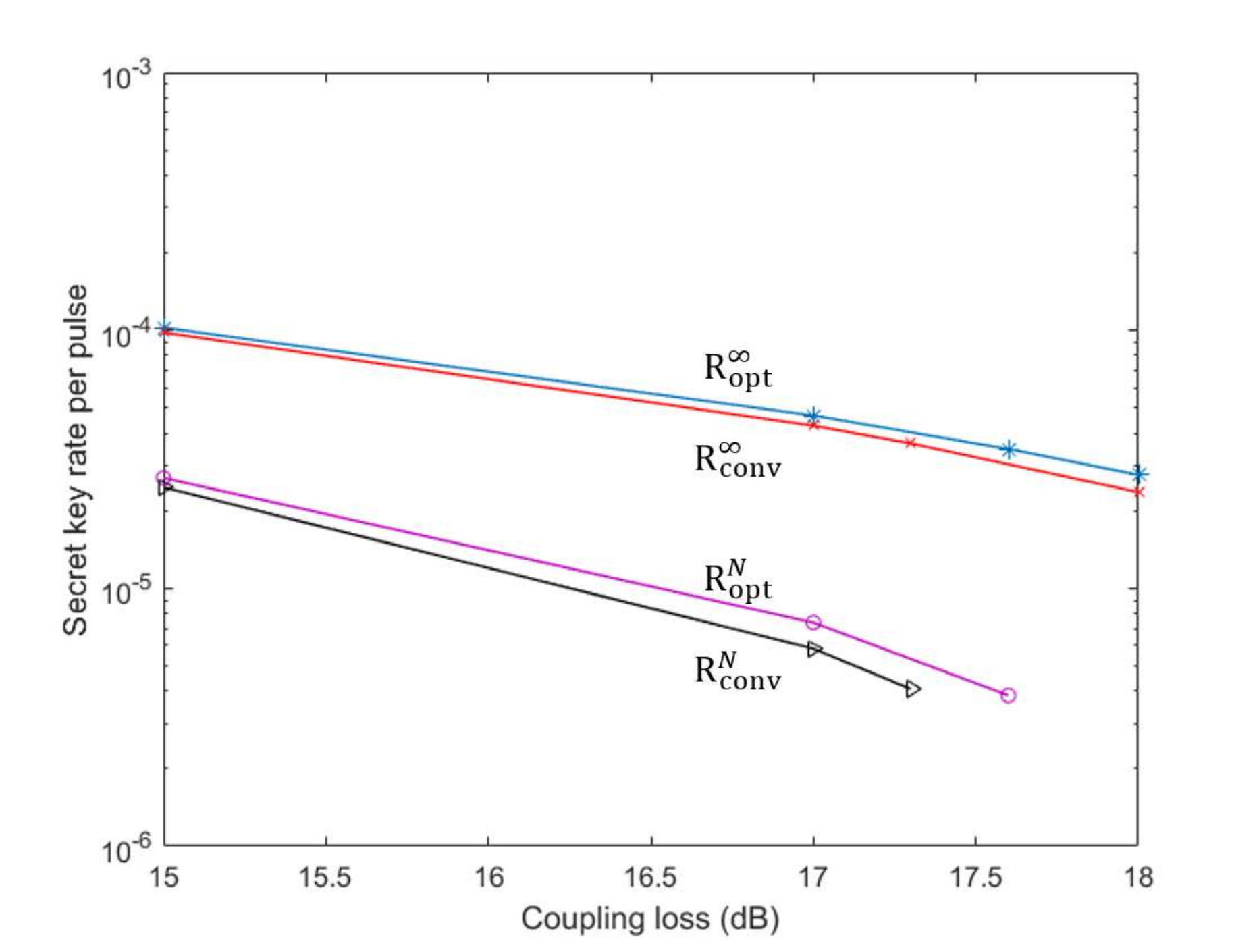}
	\caption{Average secret key rate in finite-key and asymptotic cases at different values of coupling loss for the setup of Fig.~\ref{system}(b). Here, {$P=15$}, $I=-30~{\rm dBm}$, $\Delta=0.8~{\rm nm}$, and $N=10^{11}$. {The simulations have been performed at the points represented by ``$\ast$", ``$\times$", ``$\circ$", or ``$\rhd$". The free parameters have been optimized for each individual channel at each point. As an example, at $\eta_c=17.6$, for the channel with the highest crosstalk noise in the optimal setting, we have $\mu=0.381$, $\nu=0.095$, $q^s=0.554$, $q^w=0.264$, and $P_Z=0.5$.   }   \label{fig_loss}} 
\end{figure}

Figure~\ref{fig_users} shows the average secret key rate in the wireless setup for the number of users ranging from 16 to its maximum 22. In the finite-key scenario, at $N= 10^{11}$, the conventional technique cannot support more than 21 users even though the launch power is as low as -30~dBm. This implies that in certain extreme regimes, the near-optimal technique again buys us a little bit of additional capacity. Moreover, as can be seen in all graphs so far, the rate enhancement in the finite-key regime is higher than the asymptotic case. This, as it was mentioned, is because of the sensitivity of our QBER bounds to the block size. As an example, in Fig.~\ref{fig_users}, for $P=20$, we have $\Gamma^{N}=31.48 \%$, while $\Gamma^{\infty}=10.28 \%$. 

{We have also found the cut-off launch power, i.e., the maximum launch power for which all quantum channels have a positive key rate, for different values of $P$ and $N$, which is presented in Table.~\ref{Tab:Par2}. As can be seen, the cut-off launch power decreases with the increase in the number of users, whereas it increases by using larger block sizes. For a block size of $10^{10}$ and for 15 users or more, we reach a limit that the launch power may not be sufficiently high to guarantee our target BER for the classical channels. We can resolve this issue by using a block size of $10^{11}$, which seems to be a good trade-off between the required time for key exchange and other practical aspects of the system. The improvement in the cut-off power is rather minor if we further increase the block size to $10^{12}$.}

\subsection{Coupling loss}

Another important factor in the setup of Fig.~\ref{system}(b) is the coupling loss to the fiber. As shown in Fig.~\ref{fig_loss}, at {$P=15$} and $I=-30~{\rm dBm}$, the maximum coupling loss that our system can tolerate is less than 20~dB. This is probably quite tight for the current technology as we have to collect a narrow beam of light and couple it to a single-mode fiber. But, it is not unachievable. We again observe that the near-optimal wavelength assignment can slightly improve the performance of the system, especially in the high loss case. For example, at {$\eta_{\rm c}=17~{\rm dB}$}, we obtain {$\Gamma^{N}=26 \%$} and $\Gamma^{ \infty}=9.5 \%$.

\section{Conclusions}
\label{con}
In this paper, quantum access networks based on the DWDM-PON structure were considered. In these setups, users were either directly connected to the PON, or could use wireless links in an indoor environment. We examined the possibility of secret key exchange in the finite-key regime in such systems. Various conditions and regimes of operation were considered and the average secret key rate of users was evaluated. Our numerical results showed that it would be feasible to exchange secret keys in a reasonable time of about a few minutes or less. Furthermore, a near-optimal low-complexity wavelength assignment algorithm, with the aim of optimizing the secret key rate of QKD channels, was proposed. Our numerical results showed that by applying this method, we could achieve some improvement in the key rate of our QKD channels, especially when our system had to tolerate high noises or losses. It was also concluded that the rate enhancement that can be achieved in the finite-key regime was higher than that of the asymptotic limit. While these improvements could sometime be just marginal, they could translate into whether the system was operable or not. From the running time perspective, the users would benefit an improvement on the order of 10\%--50\%, especially if a long key needs to be exchanged.  
 
\subsection*{Funding.} The authors acknowledge financial support from the UK Engineering and Physical Sciences Research Council (EPSRC) Grant No. EP/M013472/1 and the European Union H2020 Project 675662.

\subsection*{Acknowledgement.} All data generated in this paper can be reproduced by the provided methodology and equations.

\bibliographystyle{osajnl}
\bibliography{Bibli28Sept12} 

\begin{thebibliography}{10}
\newcommand{\enquote}[1]{``#1''}

\bibitem{ChinaSatQKD}
S.-K. Liao, W.-Q. Cai, W.-Y. Liu, L.~Zhang, Y.~Li, J.-G. Ren, J.~Yin, Q.~Shen,
  Y.~Cao, Z.-P. Li, F.-Z. Li, X.-W. Chen, L.-H. Sun, J.-J. Jia, J.-C. Wu, X.-J.
  Jiang, J.-F. Wang, Y.-M. Huang, Q.~Wang, Y.-L. Zhou, L.~Deng, T.~Xi, L.~Ma,
  T.~Hu, Q.~Zhang, Y.-A. Chen, N.-L. Liu, X.-B. Wang, Z.-C. Zhu, C.-Y. Lu,
  R.~Shu, C.-Z. Peng, J.-Y. Wang, and J.-W. Pan, \enquote{Satellite-to-ground
  quantum key distribution,} Nature \textbf{549}, 43 (2017).

\bibitem{ChinaSatTelep}
J.-G. Ren, P.~Xu, H.-L. Yong, L.~Zhang, S.-K. Liao, J.~Yin, W.-Y. Liu, W.-Q.
  Cai, M.~Yang, L.~Li, K.-X. Yang, X.~Han, Y.-Q. Yao, J.~Li, H.-Y. Wu, S.~Wan,
  L.~Liu, D.-Q. Liu, Y.-W. Kuang, Z.-P. He, P.~Shang, C.~Guo, R.-H. Zheng,
  K.~Tian, Z.-C. Zhu, N.-L. Liu, C.-Y. Lu, R.~Shu, Y.-A. Chen, C.-Z. Peng,
  J.-Y. Wang, and J.-W. Pan, \enquote{Ground-to-satellite quantum
  teleportation,} Nature \textbf{549}, 70 (2017).

\bibitem{secoqc}
M.~{\rm Peev \em et al.}, \enquote{The {SECOQC} quantum key distribution
  network in {Vienna},} New J. Phys. \textbf{11}, 075001 (2009).

\bibitem{QAccess_Toshiba}
B.~Fr\"ohlich, J.~F. Dynes, M.~Lucamarini, A.~W. Sharpe, Z.~Yuan, and A.~J.
  Shields, \enquote{A quantum access network,} Nature \textbf{501}, 69--72
  (2013).

\bibitem{Sasaki:TokyoQKD:2011}
M.~{\rm Sasaki \em et al.}, \enquote{Field test of quantum key distribution in
  the {Tokyo QKD Network},} Opt. Exp. \textbf{19}, 10387--10409 (2011).

\bibitem{qi2010feasibility}
B.~Qi, W.~Zhu, L.~Qian, and H.-K. Lo, \enquote{Feasibility of quantum key
  distribution through a dense wavelength division multiplexing network,} New
  Journal of Physics \textbf{12}, 103042 (2010).

\bibitem{MDInetwork}
Y.-L. Tang, H.-L. Yin, Q.~Zhao, H.~Liu, X.-X. Sun, M.-Q. Huang, W.-J. Zhang,
  S.-J. Chen, L.~Zhang, L.-X. You, Z.~Wang, Y.~Liu, C.-Y. Lu, X.~Jiang, X.~Ma,
  Q.~Zhang, T.-Y. Chen, and J.-W. Pan, \enquote{Measurement-device-independent
  quantum key distribution over untrustful metropolitan network,} Phys. Rev. X
  \textbf{6}, 011024 (2016).

\bibitem{China_BeiShang}
Q.~Zhang, F.~Xu, Y.-A. Chen, C.-Z. Peng, and J.-W. Pan, \enquote{Large scale
  quantum key distribution: challenges and solutions,} Opt. Express
  \textbf{26}, 24260--24273 (2018).

\bibitem{Telcordia_1550_1550}
N.~A. Peters, P.~Toliver, T.~E. Chapuran, R.~J. Runser, S.~R. McNown, C.~G.
  Peterson, D.~Rosenberg, N.~Dallmann, R.~J. Hughes, K.~P. McCabe, J.~E.
  Nordholt, and K.~T. Tyagi, \enquote{Dense wavelength multiplexing of 1550nm
  {QKD} with strong classical channels in reconfigurable networking
  environments,} New J. Phys. \textbf{11}, 045012 (2009).

\bibitem{Telcordia_1550_1310}
T.~E. Chapuran, P.~Toliver, N.~A. Peters, J.~Jackel, M.~S. Goodman, R.~J.
  Runser, S.~R. McNown, N.~Dallmann, R.~J. Hughes, K.~P. McCabe, J.~E.
  Nordholt, C.~G. Peterson, K.~T. Tyagi, L.~Mercer, and H.~Dardy,
  \enquote{Optical networking for quantum key distribution and quantum
  communications,} New J. Phys. \textbf{11}, 105001 (2009).

\bibitem{Shields.PRX.coexist}
K.~A. {\rm Patel \em et al.}, \enquote{Coexistence of high-bit-rate quantum key
  distribution and data on optical fiber,} Phys. Rev. X \textbf{2}, 041010
  (2012).

\bibitem{patel2014quantum}
K.~Patel, J.~Dynes, M.~Lucamarini, I.~Choi, A.~Sharpe, Z.~Yuan, R.~Penty, and
  A.~Shields, \enquote{Quantum key distribution for 10 {G}b/s dense wavelength
  division multiplexing networks,} Applied Physics Letters \textbf{104}, 051123
  (2014).

\bibitem{CVQKD-DWDM}
R.~Kumar, H.~Qin, and R.~All$\acute{\rm e}$aume, \enquote{Coexistence of
  continuous variable {QKD} with intense {DWDM} classical channels,} New
  Journal of Physics \textbf{17}, 043027 (2015).

\bibitem{HP_HandheldQKD}
J.~L. Duligall, M.~S. Godfrey, K.~A. Harrison, W.~J. Munro, and J.~G. Rarity,
  \enquote{Low cost and compact quantum key distribution,} New J. Phys.
  \textbf{8}, 249 (2006).

\bibitem{chun2017handheld}
H.~Chun, I.~Choi, G.~Faulkner, L.~Clarke, B.~Barber, G.~George, C.~Capon,
  A.~Niskanen, J.~Wabnig, D.~O'Brien, and D.~Bitauld, \enquote{Handheld free
  space quantum key distribution with dynamic motion compensation,} Opt. Exp.
  \textbf{25}, 6784--6795 (2017).

\bibitem{Elmabrok:18}
O.~Elmabrok and M.~Razavi, \enquote{Wireless quantum key distribution in indoor
  environments,} J. Opt. Soc. Am. B \textbf{35}, 197--207 (2018).

\bibitem{elmabrok2018quantum}
O.~Elmabrok, M.~Ghalaii, and M.~Razavi, \enquote{Quantum-classical access
  networks with embedded optical wireless links,} J. Opt. Soc. Am. B
  \textbf{35}, 487--499 (2018).

\bibitem{bahrani2018wavelength}
S.~Bahrani, M.~Razavi, and J.~A. Salehi, \enquote{Wavelength assignment in
  hybrid quantum-classical networks,} Sci. Rep. \textbf{8}, 3456 (2018).

\bibitem{crosstalk2016}
S.~Bahrani, M.~Razavi, and J.~Salehi, \enquote{Crosstalk reduction in hybrid
  quantum-classical networks,} Scientia Iranica D \textbf{23}, 2898--2906
  (2016).

\bibitem{Globecom2018}
S.~Bahrani, O.~Elmabrok, G.~{Curr\'as Lorenzo}, and M.~Razavi,
  \enquote{Finite-key effects in quantum access networks with wireless links,}
  in \enquote{IEEE Globecom Workshops,}  (2018).

\bibitem{OFDM-QKD}
S.~Bahrani, M.~Razavi, and J.~Salehi, \enquote{Orthogonal frequency-division
  multiplexed quantum key distribution,} Lightwave Technology, Journal of
  \textbf{33}, 4687--4698 (2015).

\bibitem{OFDM-QKD_SPIE-Photon2016}
S.~Bahrani, M.~Razavi, and J.~A. Salehi, \enquote{Orthogonal frequency division
  multiplexed quantum key distribution in the presence of {Raman} noise,} in
  \enquote{Proc. SPIE,} , vol. 9900 (2016), vol. 9900, pp. 99001C--99001C--7.

\bibitem{bahrani2016optimal}
S.~Bahrani, M.~Razavi, and J.~A. Salehi, \enquote{Optimal wavelength allocation
  in hybrid quantum-classical networks,} in \enquote{Signal Processing
  Conference (EUSIPCO), 2016 24th European,}  (Institute of Electrical and
  Electronics Engineers (IEEE), 2016).

\bibitem{curty2014finite}
M.~Curty, F.~Xu, W.~Cui, C.~C.~W. Lim, K.~Tamaki, and H.-K. Lo,
  \enquote{Finite-key analysis for measurement-device-independent quantum key
  distribution,} Nature Commun. \textbf{5}, 3732 (2014).

\bibitem{zhang2017improved}
Z.~Zhang, Q.~Zhao, M.~Razavi, and X.~Ma, \enquote{Improved key-rate bounds for
  practical decoy-state quantum-key-distribution systems,} Phys. Rev. A
  \textbf{95}, 012333 (2017).

\bibitem{Lo:EffBB84:2005}
H.-K. Lo, H.~F. Chau, and M.~Ardehali, \enquote{Efficient quantum key
  distribution scheme and a proof of its unconditional security,} Journal of
  Cryptology \textbf{18}, 133--165 (2005).

\bibitem{eraerds2010quantum}
P.~Eraerds, N.~Walenta, M.~Legre, N.~Gisin, and H.~Zbinden, \enquote{Quantum
  key distribution and 1 {G}bps data encryption over a single fibre,} New
  Journal of Physics \textbf{12}, 063027 (2010).

\bibitem{Hungarian}
J.~Munkres, \enquote{Algorithms for the assignment and transportation
  problems,} Journal of the Society for Industrial and Applied Mathematics
  \textbf{5}, 32--38 (1957).

\bibitem{GLLP_04}
D.~Gottesman, H.-K. Lo, N.~L\"utkenhaus, and J.~Preskill, \enquote{Security of
  quantum key distribution with imperfect devices,} Quant.~Inf.~Comput.
  \textbf{4}, 325 (2004).

\end{thebibliography}

\end{document}